\documentclass[prl,twocolumn,amsmath,showpacs]{revtex4}
\input {epsf.sty}
\usepackage{epsfig}
\usepackage{amsmath,amsthm,amssymb}
\begin{document}
\title{Surface Specific Heat of $^{3}$He and Andreev Bound States}
\author{H. Choi, J.P. Davis, J. Pollanen, and W. P. Halperin}
\affiliation{Department of Physics and Astronomy,\\
       Northwestern University, Evanston, Illinois 60208}

\date{Version \today}


\begin{abstract} High resolution measurements  of the specific heat of liquid
$^{3}$He   in the presence of a silver surface have been performed at
temperatures  near the superfluid transition   in the pressure range of 1 to 29
bar. The surface contribution to the heat capacity is identified with Andreev bound
states of
$^{3}$He quasiparticles that have a range of half a coherence length.
\end{abstract}

\maketitle

\vspace{11pt}

Unconventional pairing superfluids and superconductors are sensitive to
quasiparticle scattering at surfaces since all forms of scattering  are inherently
pair breaking\cite{Abr61}.  Depending on the boundary  conditions, whether 
scattering is specular or diffuse, and depending on the specific quantum state,
the  order parameter can be significantly suppressed. Correspondingly, 
quasiparticle bound states extend from a surface a distance  approximately
equal to the coherence length of the bulk superfluid.
   These states were first discussed by Andreev\cite{And64} in order to
understand the difference between charge and thermal transport at superconducting
interfaces, and they have been extensively investigated in unconventional
superconductors.  For example, the zero-bias conductance anomaly in tunneling
experiments\cite{Cov97} has been ascribed to low-energy, surface bound states and
provides a key indicator of unconventional pairing\cite{Hu94,Fog97}.  Andreev
scattering
\cite{Kur90} and Andreev bound states (ABS) are essential characteristics of 
thin superfluid films\cite{Vor03} of
$^{3}$He and they dominate the properties of superfluid  $^{3}$He contained in
the porous medium of silica aerogel\cite{Sha01,Hal05}.  In the latter case, the
ABS lead to gapless superfluidity as has been determined from their influence on
heat capacity\cite{Cho04} and thermal conductance\cite{Fis03}.  The bound states
affect physical measurements that use probes such as vibrating
wires\cite{Gue90,Bau98} for viscosity and thermometry experiments, and crystal
oscillators\cite{Ave81,Aok05} for the measurement of the acoustic impedance.

Recently Vorontsov and Sauls\cite{Vor03} have calculated the
contribution to the free energy and specific
heat of Andreev bound states in thin films of
$^{3}$He in the A-phase. For a film on a solid surface having diffuse
scattering boundary
conditions they find that there is a suppression of the superfluid
transition, $T_{c}$, as well as a substantial
reduction of the heat capacity in the superfluid state near
$T_{c}$.  Even for
thick films, where the suppression effect on the transition
temperature is negligible, the reduction of
the heat capacity near the transition can be remarkably large. In
this letter, we present measurements of
the contribution of these bound states to the heat capacity of superfluid
$^{3}$He near a silver surface close to the transition temperature.

\begin{figure}[b]
\centerline{\epsfxsize0.93\hsize\epsffile{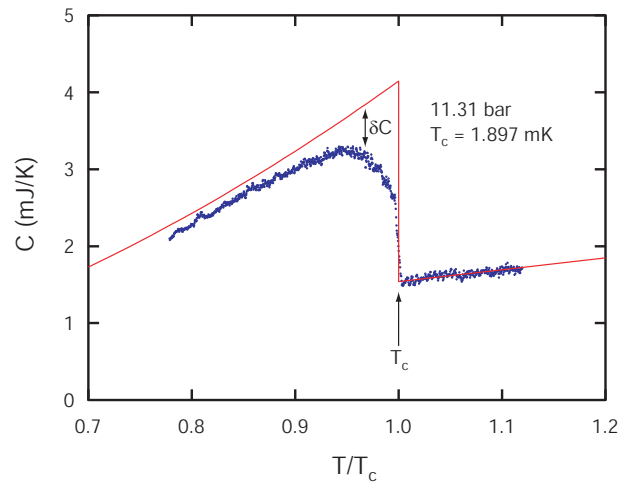}}
\begin{minipage}{0.93\hsize}
\caption {Heat capacity of both bulk and confined $^{3}$He obtained
from a slow warm up trace at 11.31 bar.
The solid trace is the heat capacity expected for bulk $^{3}$He determined from 
Greywall's measurements\cite{Gre86}. The data points are
our measurements.}
\end{minipage}
\end{figure}
\noindent

Previous experimental work on $^{3}$He in confined geometries, has
taken one of two approaches.  The first is to
investigate $^{3}$He thin layers, for example, films with a free
surface for studies of superfluid
density\cite{Xu90}, flow\cite{Sac85}, third sound\cite{Dav88}, or in
slabs having confinement on
two sides, as was the case for a number of NMR experiments\cite{Fre88}.   The
other method is to determine the effects of
surfaces on
$^{3}$He constrained in a porous medium, with the corresponding
advantage of a larger effective surface area.
If the pore structure is larger than the superfluid coherence length,
the system can be approximated as a
collection of randomly oriented planar surfaces. For measurements of the heat capacity this latter approach is preferable. Earlier experiments\cite{Gre86,Sch98,Kis00} of this
kind show that the heat capacity differs
from that of the bulk, without a consensus for  interpretation. Greywall suggested that there is  a healing
length of the superfluid at the surface\cite{Gre86}. Others have
argued\cite{Sch98,Kis00} that there is a broad
distribution of transition temperatures of disconnected superfluid
regions.  In our experiment we use a
high resolution temperature sweep method
that can provide sufficient detail to explore the temperature and
pressure dependence of the heat capacity, and we keep some
bulk helium present in the calorimeter as a reference.
Near the transition we observe a deficit in heat capacity with
respect to the pure superfluid as was found in
the thin film calculations of Vorontsov and Sauls.  The model we
develop is based on a surface specific heat
from surface Andreev bound states.  We find that the model can 
consistently account for
our results as well as those from the earlier
work.

Our measurements were performed with the calorimeter described by Choi
{\it et al.}\cite{Cho04} for high
resolution measurement of the specific heat of superfluid $^{3}$He in
silica aerogel. There are three regions
of
$^{3}$He inside the calorimeter.  The first is the interior of the silver heat exchanger 
constructed of sintered silver particles of
micron size and has a volume $V_{1}$ = $0.56 \pm 0.01$ cm$^{3}$ and
surface area  2.9 $\pm$ 0.1 m${^2}$.  The second region is the open volume
for bulk helium, $V_{2}$ =
$0.29 \pm 0.04$ cm$^{3}$.  Finally, from our earlier
studies\cite{Cho04}, we have a disk of silica aerogel
with pore volume of $V_{3}$ =
$1.06 \pm 0.01$ cm$^{3}$. The $^{3}$He in the volume $V_{3}$  remains in the
normal Fermi liquid state for all of the
experiments reported here.  We have previously determined its
volume and heat capacity to an accuracy of
2\%.  We have subtracted this contribution, plus the calorimeter
background, from our measurements and do
not discuss them further. 

The samples were cooled by adiabatic
demagnetization of PrNi$_{5}$ and the
calorimeter was isolated from this refrigerator with a
superconducting cadmium heat switch.  The temperature
of the sample cell was measured every thirty seconds using a SQUID
based mutual inductance bridge for
measurement of the magnetic susceptibility of a paramagnetic salt, La
diluted CMN.  Once the cadmium
superconducting heat switch was open, the sample cell warmed at a rate, $\dot{T}$, from an
ambient heat leak,
$\dot{Q}$, typically 0.1 nW.  Occasionally we applied external heat pulses to check
consistency and to calibrate this heat
leak.  Then the heat capacity was determined as,
\begin{equation} {C = {dQ \over dT} = {dQ \over dt} {dt \over dT} =
{\dot{Q} \over
\dot{T}}}
\end{equation}
The advantage of using slow warming traces over the adiabatic heat
pulse method is higher resolution. A heat
pulse typically causes  a temperature jump of 50 $\sim$ 100 $\mu$K.
In a slow warm-up trace, the
temperature change for each point is less than 1 $\mu$K.  However,
such a small signal inherently results in
poor signal-to-noise in determining
$\dot{T}$. This can be overcome by averaging adjacent data points
provided that the warm-up rate is adequately
slow and stable. We used averaging to smooth the data, thereby
decreasing our temperature resolution to 10
$\mu$K.  All of our slow-warming data is reproduced by our pulsed
heat capacity measurements, albeit with
lower resolution in temperature .

\begin{figure}[t]
\centerline{\epsfxsize0.93\hsize\epsffile{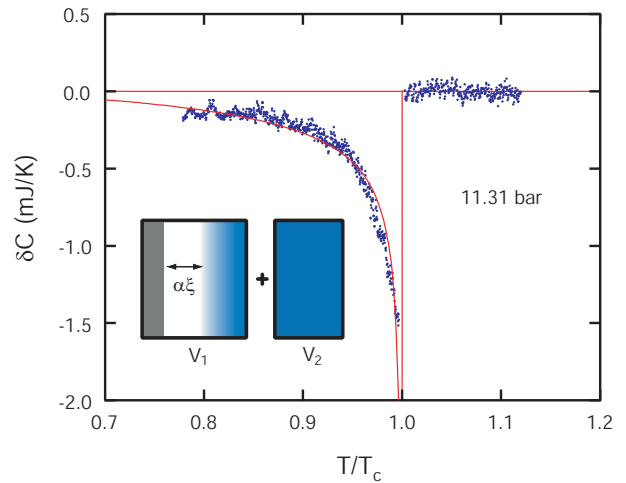}}
\begin{minipage}{0.93\hsize}
\caption {$\delta C = (C - C_{s})$ is the difference between the
measured heat capacity, $C$, and that of the
bulk superfluid, $C_{s}$, as a function of temperature at 11.31 bar.
The inset is a sketch of the volume
distribution in the calorimeter.
$V_{1}$ is the fluid inside the silver heat exchanger and $V_{2}$ the
volume outside. In our model, Andreev bound
states reside within a distance
$\alpha\xi(T,P)$ from the surface in the volume $V_{1}$; the rest of the
helium in $V_{1}$ and all of that in $V_{2}$ is taken to
be bulk superfluid. The model calculation, given by the smooth
curve with a constant scale factor $\alpha = 0.48 \pm 0.08$,  agrees well with the data.}
\end{minipage}
\end{figure}
\noindent

On cooling through $T_{c}$ we observe a sharp, resolution-limited,
increase  in the heat capacity,  shown
in Fig.1,  followed by a smooth increase and then a decrease
over a range of temperature.  For
reference we directly compare our results in this figure  with the 
heat capacity
measurements of bulk superfluid
$^{3}$He performed by Greywall\cite{Gre86}.  The central question we
address is, what is
the origin of the difference between these results.   For bulk
helium we know that the jump in heat capacity at
$T_{c}$, $\Delta C_{s}(T_{c})$, corresponds to that of a BCS pairing
system, enhanced by strong
coupling\cite{Vol90}.  The heat capacity then falls rapidly,
approximately proportional to
$T^{3}$.  Consistently, in our data we find that at
$T_{c}$ there is a sharp increase in the heat capacity and it is 
natural to identify
this jump with the bulk helium in our
calorimeter.  In Fig.2 we show the difference between the measured heat 
capacity and
that of the bulk superfluid for the same volume, $\delta
C = C - C_{s}$  as a function of temperature.  The magnitude of the 
discontinuity in
$\delta C$ at
$T_{c}$  corresponds to the amount of helium in
the silver heat exchanger given by the volume ratio
$V_{1}/(V_{1}+V_{2}) = \delta C(T_{c})/\Delta C_{s}(T_{c})$
and is plotted in Fig.3.  The  apparent volume $V_{1}$, deduced in
this way, is $0.40 \pm 0.02$ cm$^{3}$.   As expected,
it does not vary with pressure. The magnitude of the apparent volume is
qualitatively consistent with an independent measurement,
$V_{1} = 0.56 \pm 0.01$ cm$^{3}$.  Apart from experimental uncertainty this discrepancy
reflects difficulty in making an accurate extrapolation to $T_{c}$, which we discuss in
greater detail below.

\begin{figure}[b]
\centerline{\epsfxsize0.93\hsize\epsffile{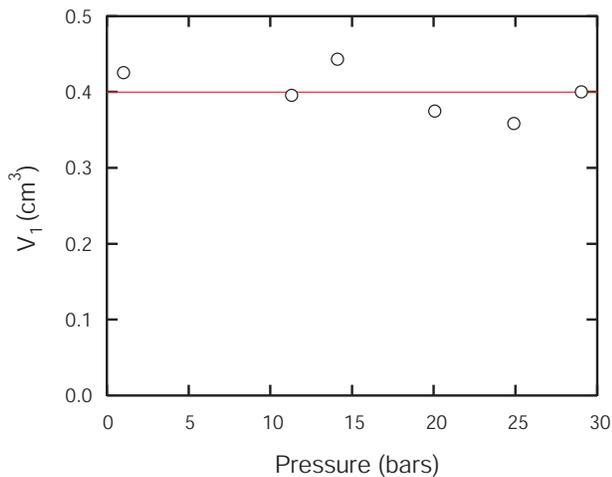}}
\begin{minipage}{0.93\hsize}
\caption {Measurements of the heat capacity discontinuity at $T_{c}$
interpreted as the volume in the heat
exchanger, $V_{1}$ for various pressures. }
\end{minipage}
\end{figure}
\noindent

\begin{figure}[t]
\centerline{\epsfxsize0.93\hsize\epsffile{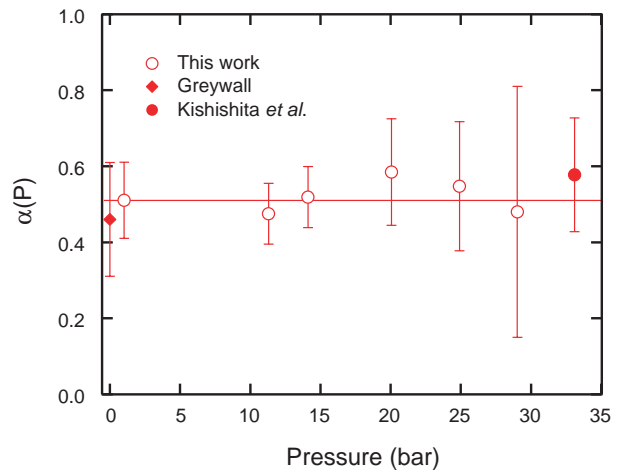}}
\begin{minipage}{0.93\hsize}
\caption {The scale factor $\alpha$ for the surface heat capacity
as a function of pressure.
The present measurements are shown
as open circles and are pressure independent with an average of $\alpha = 0.51$. There is good agreement with
earlier work from Greywall\cite{Gre86}(diamond) and Kishishita
{\it et al.}\cite{Kis00} (solid circle) interpreted in terms of  our model. }
\end{minipage}
\end{figure}
\noindent

Below $T_{c}$ the behavior of the heat capacity
must be attributed to the combination of the
surface  dominated heat capacity in the silver heat
exchanger  in addition to that of the bulk.
The formation of surface bound states corresponds to transfer of
spectral weight from above the energy gap to
low energy (near the Fermi energy) with a density of states of low energy excitations
of order that in normal helium.  Their spatial extent from the surface is
expected\cite{Kja78,Vor03} to be approximately that of the
coherence length given by
$\xi(T,P) = \xi_{0}(P)(1 - T/T_{c})^{-1/2}$
where
$\xi_{0}(P) = \hbar v_{F}/ 2\pi k_{B}T_{c}$. Here $v_{F}$ is the Fermi
velocity and $k_{B}$ is the Boltzman constant.
On this basis we propose a simple model where we 
take the surface heat capacity to be
proportional to that of the normal fluid but constrained to a volume
that scales as $A\xi(T,P)$ where $A$ is the
area of the silver surface. Consequently, we write the surface
contribution to the heat capacity as
$\alpha(T,P)\xi(T,P)Ac_{n}$ and we investigate the temperature and pressure dependence of the scale factor
$\alpha$.  A pictorial representation of this model is 
sketched in
the inset of Fig.2.  The corresponding heat capacity is,
\begin{align} C &= \alpha(T,P)\xi(T,P)Ac_{n} +
(V_{1}+V_{2}-\alpha(T,P)\xi(T,P)A)c_{s}\notag \\
&=(V_{1}+V_{2})c_{s}+\alpha(T,P)\xi(T)A(c_{n}-c_{s})\quad
\end{align} and $\delta C = \alpha(T,P)\xi(T,P)A(c_{n}-c_{s})$.  The
scale factor, $\alpha(T,P)$ is the
only unknown parameter necessary to describe the surface
heat capacity and expresses the temperature
dependence and pressure dependence of the surface heat capacity
beyond that given by the coherence length and the normal specific heat.

We have used this model to interpret our measurements for various temperatures
and  pressures.  For any given pressure, we find $\alpha$ is constant over
the  available range of temperature, down to
$T/T_{c} \approx 0.7$.  In Fig.2. our measurements of $\delta C$
at a pressure of 11.31 bar are compared with a fit to Eq.2 taking $\alpha$ to be
temperature independent.  The good agreement between the data and the calculated curve 
confirms that 
$\alpha $ is a constant with a best fit value of $0.48\pm0.08$ at
this pressure.   The significant down turn in $\delta C$ near $T_{c}$ in Fig.2 is due to the strong
temperature dependence of the coherence length.  We have made this comparison at all pressures and the
results are presented together in Fig.4.  The scale factor appears to be both  temperature {\it and}
pressure  independent  with the  average value  
$\alpha =0.51\pm 0.15$.  In the context of our model this means that the
spatial range for surface excitations, that we associate with quasiparticle bound
states, is  a half of a coherence length.

Greywall\cite{Gre86} allowed for a healing length of superfluid
$^{3}$He near the silver heat exchanger
surface in his measurement of specific heat and he assumed its temperature
dependence to have the form
$(1-(T/T_{c})^{4})^{-1/2}$. We have reanalyzed his data with
our model as well as the work of Kishishita {\it et al.}\cite{Kis00}. 
Both results are plotted in Fig.4 where they are
compared directly to ours. 
It is noteworthy that the silver sinter used by Kishishita {\it et al.} had an area-to-volume ratio of
$12\times10^{6}$ m$^{-1}$, the one in the Greywall experiment was
$3.4\times10^{6}$ m$^{-1}$, and these can be
compared with ours,
$5.2\times10^{6}$ m$^{-1}$. There is excellent  agreement among the
experiments, although they are
performed in a range of pore structures with area-to-volume ratios
spaning a factor of three.  This implies
that different structures among the silver sinters do not 
play a role.
The overall consistency of the data with the model, including the variables of
pressure,  temperature, and different silver surface structures, provides
compelling evidence that we are measuring a surface contribution to
the heat capacity, rather than the heat
capacity of disconnected regions of superfluid with a distribution of
transition temperatures.  However, the
model will not be correct close to the transition temperature where
the coherence length diverges with
increasing temperature approaching $T_{c}$.  There is a point,
nominally a few percent lower than
$T_{c}$, where $\alpha\xi(T,P)=\alpha\xi_{0}(P)(1-T/T_{c})^{-1/2}$ reaches
$V_{1}/A$ = 193 nm.  In our model,
$\alpha\xi(T,P)A$ represents the volume of the surface bound states
and, at this temperature, they would fill
the silver exchanger of volume $V_{1}$. At 1.024 bars this excluded
temperature region is 
3\% of $T_{c}$; it decreases with decreasing coherence length at
higher pressure. Additionally, there is a
small  suppression of the transition temperature for helium in
restricted geometry. These effects may
account for difficulty in extending our model close to $T_{c}$ and
the corresponding systematic error from
such an extrapolation in determining $V_{1}$ as is shown in Fig.3.

From a theoretical perspective, quasiparticle scattering at the
surface is responsible for a
non-zero density of states at the Fermi level which should give a
heat capacity that is linear in
the temperature in the low temperature limit.  Our model for the surface
specific heat has this temperature dependence at low temperatures where 
$c_{s} = 0$ in Eq.2.  Additionally, the entropy  at the transition temperature 
determined from the specific heat in the
model is within a few percent of that of the normal fluid at $T_{c}$, as is 
required for a second order thermodynamic transition.  
Although Eq.2 is highly phenomenological, nonetheless  it might be a useful guide over
a wider range of temperature than we have explored.  It gives a low temperature limit 
for the density of states, relative
to the normal fluid, to be simply proportional to the pressure dependent coherence
length,
$\alpha \xi_{0}(P) A/V_{1}$. It would be interesting to extend heat capacity experiments 
to lower temperatures  for a direct measurement of the density of states of surface
bound states.  

In conclusion, we have used a high resolution method to determine the
heat capacity of $^{3}$He in the
presence of a silver surface.  We distinguish two different
contributions; one from the bulk superfluid phase
and the other from the helium near the silver surface. We have
constructed a model based on low energy
contributions to the density of states associated with Andreev bound
states of $^{3}$He quasiparticles that
scatter from the surface.  We have found that the surface heat capacity has a
temperature and pressure dependence given  by the normal fluid specific
heat and the bulk  $^{3}$He coherence  length.  Further, we determine 
that the spatial extent of
the bound state region is one  half of the bulk  $^{3}$He 
coherence  length.  Our
confirmation of the existence of surface bound states from 
measurement of the heat  capacity
supports recent  results from surface sensitive measurements of the
transverse acoustic impedance by Aoki {\it et
al.}\cite{Aok05} and theoretical calculations of superfluid $^{3}$He in slabs by
Vorontsov and Sauls\cite{Vor03}.

We acknowledge support from the National Science Foundation, DMR-0244099
and helpful discussions with Jim Sauls and Anton Vorontsov.

\end{document}